\documentstyle[12pt]{article}\textheight 230mm\textwidth 160mm
\newcommand{\bi}{\bibitem}
\newcommand{\be}{\begin{eqnarray}}
\newcommand{\ee}{\end{eqnarray}}
\newcommand{\nn}{\nonumber}

\begin{document}
\hspace*{11cm}\vspace{-2mm}MPI-Ph/94-55\\
\hspace*{11.6cm}\vspace{-2mm}HD-THEP-94-34\\
\hspace*{11.6cm}\vspace{-2mm}KANAZAWA-94-18\\
\hspace*{11.6cm}September 1994

\vspace{1 cm}
\begin{center}
 {\Large\bf Finite Unification and Top Quark Mass $^{\dag}$}
\end{center}

\vspace*{0.1cm}
\begin{center}{\sc Jisuke Kubo}$\ ^{(1),*}$,
{\sc Myriam
Mondrag{\' o}n}$\ ^{(2)}$\vspace{-1mm} \\
{\sc George Zoupanos}$\ ^{(3),**}$
\end{center}
%\vspace*{0.1cm}
\begin{center}
{\em $\ ^{(1)}$ Max-Planck-Institut f\"ur Physik,
 Werner-Heisenberg-Institut \vspace{-2mm}\\
D-80805 Munich, Germany} \\
{\em $\ ^{(2)}$ Institut f{\" u}r Theoretische Physik,
Philosophenweg 16\vspace{-2mm}\\
D-69120 Heidelberg, Germany}\\
{\em $\ ^{(3)}$ Physics Department, National Technical\vspace{-2mm}
University\\ GR-157 80 Zografou, Athens, Greece }  \end{center}

\noindent
{\sc\large Abstract}

\noindent
In unified gauge theories there
exist renormalization group invariant
relations among gauge and Yukawa couplings that are compatible
with perturbative renormalizability, which could be considered as a
Gauge-Yukawa Unification. Such relations are even necessary  to ensure
all-loop finiteness in Finite Unified Theories, which have vanishing
$\beta$-functions beyond the unification point.
We elucidate this alternative way of unification, and then
present its phenomenological consequences in
$SU(5)$-based models.

\vspace*{3cm}
\footnoterule
\vspace*{2mm}
\noindent
$^{*}$On leave of absence from
 College of Liberal Arts, Kanazawa Univ., Japan\vspace{-2mm}\\
$^{**}$Partially supported by C.E.U. projects
(SC1-CT91-0729;CHRX-CT93-0319).\\
$ ^{\dag}$ Presented by G. Zoupanos at the  \vspace{-2mm}
{\em 27th International Conference on High
Energy Physics},\\
Glasgow, 20-27th July 1994, to appear in the
proceedings.

\newpage
\pagestyle{plain}
\section{Introduction}
The original unification philosophy \cite{georgi1,fritzsch1}
relates the gauge and separately
the Yukawa couplings. A logical extension is to relate
the couplings of the two sectors; Gauge-Yukawa
Unification (GYU).
Within the assumption that all the particles appearing in a
theory are elementary,
the theories based on extended supersymmetries
\cite{fayet1} and string theories \cite{string}
are well-known possibilities for GYU.
Unfortunately, these theories
seem to introduce more
serious and difficult phenomenological problems to be solved than those of
the standard model.

There exists an alternative way to
unify couplings
which is based on the fact that
within the framework of
renormalizable field theory, one can find renormalization group
invariant (RGI) relations among parameters
and improve in this way the calculability
and predictive power of a given
theory \cite{zimmermann1}-\cite{kubo2}.
We would like to briefly outline this idea below.
Any RGI relation among couplings
(which does not depend on the renormalization
scale $\mu$ explicitly) can be expressed,
in the implicit form $\Phi (g_1,\cdots,g_A) ~=~\mbox{const.}$,
which
has to satisfy the partial differential equation (PDE)
\be
\mu\,\frac{d \Phi}{d \mu} &=& {\vec \nabla}\cdot {\vec \beta} ~=~
\sum_{a=1}^{A}
\,\beta_{a}\,\frac{\partial \Phi}{\partial g_{a}}~=~0~,
\ee
where $\beta_a$ is the $\beta$-function of $g_a$.
This PDE is equivalent
to the set  to ordinary differential equations,
the so-called reduction equations (REs)\cite{zimmermann1},
\be
\beta_{g} \,\frac{d g_{a}}{d g} &=&\beta_{a}~,~a=1,\cdots,A~,
\ee
where $g$ and $\beta_{g}$ are the primary
coupling and its $\beta$-function,
and $a$ does not include it.
Since maximally ($A-1$) independent
RGI ``constraints''
in the $A$-dimensional space of couplings
can be imposed by $\Phi_a$'s, one could in principle
express all the couplings in terms of
a single coupling $g$.
 The strongest requirement is to demand
 power series solutions to the REs,
\be
g_{a} &=& \sum_{n=0}\rho_{a}^{(n+1)}\,g^{2n+1}~,
\ee
which formally preserve perturbative renormalizability.
Remarkably, the
uniqueness of such power series solutions
can be decided already at the one-loop level \cite{zimmermann1}.
To illustrate this, let us assume that the $\beta$-functions
have the form
\be
\beta_{a} &=&\frac{1}{16 \pi^2}[
\sum_{b,c,d\neq g}\beta^{(1)\,bcd}_{a}g_b g_c g_d+
\sum_{b\neq g}\beta^{(1)\,b}_{a}g_b g^2]+\cdots~,\nn\\
\beta_{g} &=&\frac{1}{16 \pi^2}\beta^{(1)}_{g}g^3+
\cdots~,
\ee
where $\cdots$ stands for higher order terms, and
$ \beta^{(1)\,bcd}_{a}$'s are symmetric in $
b,c,d$.
 We then assume that
the $\rho_{a}^{(n)}$'s with $n\leq r$
have been uniquely determined. To obtain $\rho_{a}^{(r+1)}$'s,
we insert the power series (3) into the REs (2) and collect terms of
$O(g^{2r+3})$ and find
\be
\sum_{d\neq g}M(r)_{a}^{d}\,\rho_{d}^{(r+1)} &=&
\mbox{lower order quantities}~,\nn
\ee
where the r.h.s. is known by assumption, and
\be
M(r)_{a}^{d} &=&3\sum_{b,c\neq g}\,\beta^{(1)\,bcd}_{a}\,\rho_{b}^{(1)}\,
\rho_{c}^{(1)}+\beta^{(1)\,d}_{a}
-(2r+1)\,\beta^{(1)}_{g}\,\delta_{a}^{d}~,\\
0 &=&\sum_{b,c,d\neq g}\,\beta^{(1)\,bcd}_{a}\,
\rho_{b}^{(1)}\,\rho_{c}^{(1)}\,\rho_{d}^{(1)}
+\sum_{d\neq g}\beta^{(1)\,d}_{a}\,\rho_{d}^{(1)}
-\beta^{(1)}_{g}\,\rho_{a}^{(1)}~.
\ee
 Therefore,
the $\rho_{a}^{(n)}$'s for all $n > 1$
for a given set of $\rho_{a}^{(1)}$'s can be uniquely determined if
$\det M(n)_{a}^{d} \neq 0$  for all $n \geq 0$.

The possibility of coupling unification described above
is without any doubt
attractive because the ``completely reduced'' theory contains
only one independent coupling, but  it can be
unrealistic. Therefore, one often would like to impose fewer RGI
constraints, and this is the idea of partial reduction \cite{kubo1}.
Among the existing possibilities
in the framework of susy $SU(5)$ GUTs,
there are two models that are singled out by being strongly
motivated \cite{kapet1,kubo2}. The first is the $SU(5)$-Finite
Unified Theory (FUT) \cite{kapet1}.
In this theory, there exist
RGI relations among gauge and Yukawa couplings that yield
the vanishing of all $\beta$-functions
to all orders in perturbation theory \cite{sibold1}.
(It has been recently found that the quantum corrections to
the cosmological constant in a finite theory is
weakened \cite{cos}.)
 The second is the minimal $SU(5)$
susy model which can be successfully partially-reduced \cite{kubo2}.
This model is
attractive because of its simplicity. In the
following, we will give more emphasis in
discussing the $SU(5)$-FUT and then we
compare the predictions of the two models.

\section{$N=1$ Finiteness}
  Let us consider a chiral, but anomaly free,
globally supersymmetric
gauge theory based on a simple group $G$ with the gauge coupling
$g$. The
superpotential of the theory is given by
\be
W&=& \sum_{i,j}\frac{1}{2}\,m_{ij} \,\phi^{i}\,\phi^{j}+
\frac{1}{6}\sum_{i,j,k}\,\lambda_{ijk}
\,\phi^{i}\,\phi^{j}\,\phi^{k}~,
\ee
where
the matter chiral superfield $\phi^{i}$
belongs to an irreducible representation
of $G$.
The non-renormalization theorem ensures that
there are no extra mass
and cubic-interaction-term renormalizations, implying that
the $\beta$-functions of
$\lambda_{ijk}$ can be expressed as linear combinations of the
anomalous dimension matrix $\gamma_{ij}$ of
$\phi^{i}$.
Therefore, all the one-loop $\beta$-functions of the theory vanish
if
\be
\beta_{g}^{(1)}&=&0~\mbox{and}~\gamma_{ij}^{(1)}~=~0~
\ee
are satisfied,
where $\beta_{g}^{(1)}$  and $\gamma_{ij}^{(1)}$
are the one-loop coefficients
of $\beta_g$ and $\gamma_{ij}$, respectively.
A very interesting result is that these conditions (8) are
necessary and sufficient for finiteness at
the two-loop level \cite{finite}.

A natural question is what happens in higher loops.
Since the finiteness conditions impose relations
among couplings, they have to be consistent with the
REs (1).
(This should be so even for the one-loop finiteness.)
Interestingly,
 there exists a powerful theorem \cite{sibold1}
which provides the necessary and sufficient conditions for
finiteness to all loops.
The theorem makes heavy use of the non-renormalization
property of the supercurrent anomaly \cite{sibold2}.
In fact, the  finiteness theorem can be formulated in terms of
one-loop quantities, and it states
that for susy gauge theories we are considering here, the necessary
and sufficient conditions for $\beta_{g}$ and $\beta_{ijk}$ to
vanish to all orders are \cite{sibold1}: \newline
(a) The validity of the one-loop finiteness conditions, i.e.,
 eq. (8) is satisfied.
\newline
(b) The REs (2) admit a unique power series
solution, i.e., the corresponding matrix $M$ defined in eq. (5) with
$\beta_{g}^{(1)}=0$ has to be non-singular.
\newline
The latter condition
is equivalent to the requirement that the one-loop
solutions
$\rho_{a}^{(1)}$'s
are isolated and non-degenerate. Then each of these solutions
can be extended, by a recursion formula, to a formal power series in $g$
giving a theory which depends on a single coupling $g$, and has
$\beta$-functions vanishing to all orders.

\section{Finite Unified Models based on $SU(5)$}

{}From the classification of
theories with $\beta_{g}^{(1)}=0$
\cite{finite2}, one can see that
using $SU(5)$ as gauge group there
exist only two candidate models which can accommodate three fermion
generations. These models contain the chiral supermutiplets
${\bf 5}~,~\overline{\bf 5}~,~{\bf 10}~,~\overline{\bf 5}~,~{\bf 24}$
with the multiplicities $(6,9,4,1,0)$ and
 $(4,7,3,0,1)$, respectively.
Only the second one contains a ${\bf 24}$-plet which can be used
for spontaneous symmetry breaking (SSB) of $SU(5)$ down
to $SU(3)\times SU(2) \times U(1)$. (For the first model
one has to incorporate another way, such as the Wilson flux
breaking to achieve the desired SSB of $SU(5)$.)
Here we would like to concentrate only on the second model.

The most general $SU(5)$ invariant, cubic
superpotential of the (second)
model is:
\be
W &=&H_{a}\,[~
f_{ab}\,\overline{H}_b {\bf 24}+
h_{ia}\,\overline{\bf 5}_i {\bf 24}
+\overline{g}_{ija}\,{\bf 10}_i \overline{\bf 5}_{j}]+
 p\,({\bf 24})^3 \nn\\
&+& \frac{1}{2}\,{\bf 10}_i\,[~
g_{ija}\,{\bf 10}_j H_a+
 \hat{g}_{iab}\,\overline{H}_a
\overline{H}_b+
g_{ijk}^{\prime}\,
\overline{\bf 5}_{j} \overline{\bf 5}_{k}~]~,
\ee
where $i,j,k=1,2,3$ and $a,b=1,\cdots,4$, and we sum over all
indices  in $W$ (the $SU(5)$ indices are suppressed).
The ${\bf 10}_{i}$'s
and $\overline{\bf 5}_{i}$'s are the usual
three generations, and the four
$({\bf 5}+ \overline{\bf 5})$ Higgses are denoted by
 $H_a~,~\overline{H}_{a} $.

Given the superpotential, the
$\gamma^{(1)}$'s can be easily computed ($\beta_{g}^{(1)}$
vanishes of course). To ensure finiteness of the model
to all orders, we have to find
$\rho^{(1)}$'s that are isolated and non-degenerate solutions
of eq. (6) and
are consistent with the vanishing $\gamma^{(1)}$'s.
In most of the previous studies of
the present model \cite{finite3}, however,
no attempt was made to find isolated and non-degenerate
solutions, but rather the opposite. They have used the freedom
offered by the degeneracy in order to make specific ans{\" a}tze
that could lead to phenomenologically acceptable
predictions (see also \cite{finite4}). Here we concentrate on finding an
isolated and non-degenerate solution that is phenomenologically
interesting. As a first approximation to the Yukawa
matrices, a diagonal solution, that is, without
intergenerational mixing, may be considered.
It has turned out that this can be achieved by imposing
the $Z_{7}\times Z_{3}$ discrete symmetry and
a multiplicative $Q$-parity  on $W$, and that, in order
to respect these symmetries,
only $ g_{iii}~,~\overline{g}_{iii}~,
{}~f_{ii}$ and $p$ are allowed to be non-vanishing.
Moreover, we have found that under this situation
there exists a unique reduction solution that satisfies the finiteness
conditions (a) and (b) \cite{kapet1}:
\be
g_{iii}^{2}&=&\frac{8}{5}g^2+O(g^4)~,~
\overline{g}_{iii}^{2}~=~\frac{6}{5}g^2+O(g^4)~,~
f_{ii}~=~0~,\nn\\
f_{44}^{2}&=&g^2+O(g^4)~,~p^2~=~\frac{15}{7}g^2+O(g^4)~,
\ee
where $i=1,2,3$, and
the $O(g^4)$ terms are power series in $g$ that can be uniquely
computed to any finite order if the $\beta$-functions
of the unreduced model are known to the corresponding order.
The reduced model in which gauge and Yukawa couplings
are unified
has the $\beta$-functions that identically vanish to that order.

\section{Phenomenological Consequences}
In the above model, we found a diagonal solution for the Yukawa
couplings, with each family coupled to a different Higgs.
However, we may use the fact that mass terms
do not influence the $\beta$-functions in a certain
class of renormalization schemes, and introduce
appropriate mass terms that permit us to perform a rotation in the Higgs
sector such that only one pair of Higgs doublets, coupled to
the third family, remains light and acquires a
non-vanishing v.e.v.
Note that the effective coupling of the Higgs doublets
to the first family after
the rotation is very small avoiding in this way a potential problem
with the proton lifetime \cite{proton}.
Thus, effectively,
we have at low energies the minimal susy standard model with
only one pair of Higgs doublets. Adding soft
breaking terms (which are supposed not to influence the
$\beta$-functions beyond $M_{GUT}$),
we can obtain susy breaking.
The conditions on the soft breaking terms to reserve
one-loop finiteness are given in \cite{soft}. Recently, the same problem
at the two-loop level has been addressed \cite{jones}.
It is an open problem whether there exists a suitable set of conditions
on the soft terms for all-loop finiteness.
Since the $SU(5)$ symmetry is spontaneously broken
below $M_{GUT}$, the finiteness conditions obviously
do not restrict the renormalization property at low energies, and
all it remains is a boundary condition on the
gauge and Yukawa couplings; these couplings at low energies
have to be so chosen that they satisfy  (10) at $M_{GUT}$.
So we examine the evolution of the gauge couplings according
to their renormalization group equations at two-loops.
Representative results are summarized in table 1.
(To simplify our numerical analysis  we assume a unique
threshold $M_{S}$ for all the  superpartners.)

\vspace{0.3 cm}
\hspace{1.5 cm}
\begin{tabular}{|c|c|c|c|c|c|c|}
\hline
$M_{S}$ [TeV]& $\alpha_{S}(M_{Z})$ &
 $\tan \beta$  & $m_{b}$ [GeV]& $m_{t} [GeV]$
\\ \hline
$1.0$ &$0.117$ & $54.1$ &
 $5.13$ & $185$
\\ \hline
$0.5$   & $0.121$ & $53.5$
 & $5.27$
 & $186$  \\ \hline
$0.2$  & $0.121$ &  $54.1$
 & $5.14$
 & $185$  \\ \hline
\end{tabular}
\begin{center}
{\bf Table 1}. The predictions
for different $M_{S}$, where we have used:\\
$m_{\tau}=1.78$ GeV,
$\alpha_{em}^{-1}(M_{Z})=127.9$ and $\sin\theta_{W}(M_{Z})
=0.232$.
\end{center}
All the quantities except $M_{S}$ in table 1 are predicted.
The dimensionless parameters
(except $\tan \beta$) are defined in the $\overline{\rm MS}$
scheme, and the masses are pole masses.
We see from table 1 that the low energy predictions
are relatively stable against  the change of$M_{S}$ and
$m_{t}$ agrees with the CDF result \cite{cdf1}.

To compare the predictions above with those of the partially-reduced,
minimal $SU(5)$ GUT \cite{kubo2}, we present
its predictions in table 2.

\vspace{0.3 cm}
\hspace{1.5 cm}
\begin{tabular}{|c|c|c|c|c|c|c|c|}
\hline
$M_{S}$ [TeV]&$\alpha_{S}(M_{Z})$ &
$\tan \beta$  & $m_{b}$ [GeV]& $m_{t} [GeV]$
\\ \hline
$1.0$ &$0.118$ & $47.4$ &
 $5.36$ & $180$
\\ \hline
$0.5$   & $0.120$ & $47.6$ &
  $5.42$
 & $180$  \\ \hline
$0.2$  & $0.124$ & $47.4$
 & $5.55$
 & $182$  \\ \hline
\end{tabular}
\begin{center}
{\bf Table 2}. The predictions of the partially-reduced,
minimal $SU(5)$ GUT for the same low-energy inputs.
\end{center}
We see from table 1 and 2 that
the predictions of the partially-reduced,
minimal $SU(5)$ GUT do not
differ very much from these of the
$SU(5)$-FUT model.

We would like to stress that both models have the strongest
predictive power as compared with any other known GUTs
as it was promised.

\vspace{0.3 cm}
\noindent
We would like to thank  D. Matalliotakis, L. Nellen,
R. Oehme, K. Sibold, T. Yanagida  and W. Zimmermann for useful
discussions.

\end{document}